# Non-Destructive Low-Temperature Contacts to MoS$_2$ Nanoribbon and Nanotube Quantum Dots

*Robin T. K. Schock, Jonathan Neuwald, Wolfgang Möckel, Matthias Kronseder, Luka Pirker, Maja Remškar, and Andreas K. Hüttel\**

Molybdenum disulfide nanoribbons and nanotubes are quasi-1D semiconductors with strong spin–orbit interaction, a nanomaterial highly promising for quantum electronic applications. Here, it is demonstrated that a bismuth semimetal layer between the contact metal and this nanomaterial strongly improves the properties of the contacts. Two-point resistances on the order of 100 kΩ are observed at room temperature. At cryogenic temperature, Coulomb blockade is visible. The resulting stability diagrams indicate a marked absence of trap states at the contacts and the corresponding disorder, compared to previous devices that use low-work-function metals as contacts. Single-level quantum transport is observed at temperatures below 100 mK.

## 1. Introduction

Following the isolation of graphene,[1] many quasi-planar 2D materials were discovered and tested for their electronic properties,[2,3] including in particular the transition metal dichalcogenides[4] (TMDCs) MX$_2$ with M = W, Mo, ... and X = S, Se, Te. Molybdenum disulfide MoS$_2$, one of these compounds, is a semiconductor with strong spin–orbit interaction, and displays intrinsic superconductivity at strong n-doping.[5–8] While much effort has been invested in building quantum dot devices from planar TMDCs, the effective electron mass in the conduction band[9–11] places stringent requirements on quantum confinement. As a result, lithographically defined devices so far demonstrate classical Coulomb blockade,[12–23] and observations of quantum effects are, with the notable exception of one recent publication[24] on WSe$_2$, limited to uncontrolled trap states.[14,25,26]

Quasi-1D TMDC nanoribbons and nanotubes[27,28] with intrinsic narrow geometric confinement can here clearly provide larger quantization energies. So far the synthesis of single-wall MoS$_2$ nanotubes still poses challenges.[29,30] MoS$_2$ multiwall nanotubes and nanoribbons, however, can be grown long, straight, and with a low defect density.[28,31] Disordered Coulomb blockade has already been observed, and even tentative indications of quantum excitations have been detected.[32,33]

A central challenge now is to obtain charge-trap-free, non-destructive, and ideally near-Ohmic contacts to these nanomaterials. Typically a metal–MoS$_2$ contact leads to Fermi level pinning close to the conduction band edge.[34] For minimizing the Schottky barrier, this suggests the use of low-work function contact metals, and indeed best results have been observed so far with titanium or scandium.[32,35,36] Since these metals have a larger affinity to sulphur than molybdenum, surface deposition alone is however sufficient to destroy the MoS$_2$ lattice several layers deep.[37] Specifically in low-temperature applications this leads to strong disorder and charge trapping below the contacts.[32]

Here, we demonstrate transparent conduction and semi-regular Coulomb blockade, tuneable via an applied gate voltage, even in ultralow-temperature ($T_{\text{base}} \simeq 15\,\text{mK}$) experiments. This is based on the recent discovery that room temperature Ohmic contacts to planar MoS$_2$ can be achieved using the semimetal bismuth:[38] since Fermi level pinning is caused by a hybridization of metal and semiconductor states at the interface,[39] a reduction of the contact density of states near the Fermi energy counterintuitively allows for tunability and transparent conduction. While (likely substrate-induced) disorder is still present, our data indicate the marked absence of charge traps at the contacts, and low contact resistances. This represents a significant improvement of contact quality. In the low-temperature limit $T \lesssim 100\,\text{mK}$, we observe indications of single-level transport.

## 2. Material and Device Fabrication

The MoS$_2$ nanomaterial, as shown in **Figure 1**a, is grown via a slow chemical transport reaction using iodine as a transport

R. T. K. Schock, J. Neuwald, W. Möckel, M. Kronseder, A. K. Hüttel
Institute for Experimental and Applied Physics
University of Regensburg
93040 Regensburg, Germany
E-mail: andreas.huettel@physik.uni-regensburg.de

L. Pirker, M. Remškar
Solid State Physics Department
Jožef Stefan Institute
1000 Ljubljana, Slovenia

L. Pirker
J. Heyrovský Institute of Physical Chemistry, v.v.i.
Czech Academy of Sciences
182 23 Prague, Czech Republic

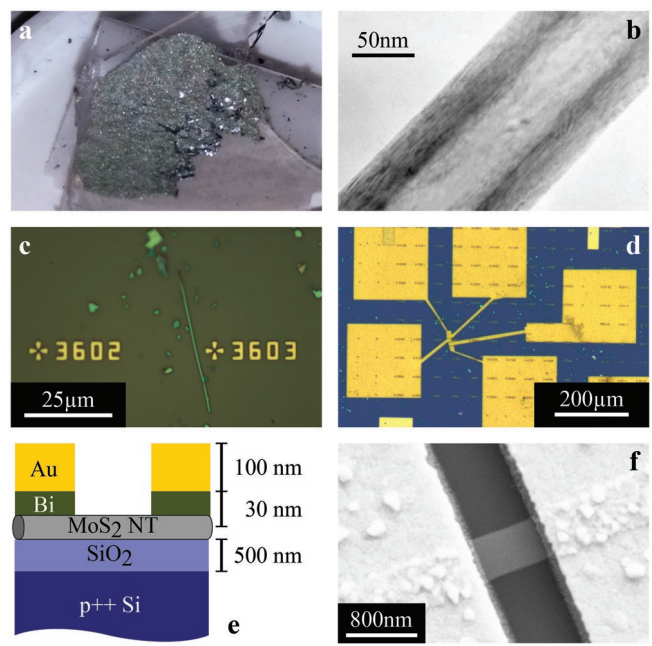

**Figure 1.** a) As-grown MoS$_2$ nanomaterial. b) Example transmission electron microscopy (TEM) image of a MoS$_2$ nanotube with a diameter of $d = 118$ nm. c) Optical microscopy image of a MoS$_2$ nanoribbon deposited on a Si/SiO$_2$ substrate. d) Finished device with contact electrodes and bond pads. e) Schematic of the cross-sectional material layers. f) SEM image of an active device region with a nanoribbon.

agent.[28,31] This results in MoS$_2$ flakes as well as nanoribbons and nanotubes of a wide range of diameter, see, e.g., Figure 1. Using a wafer dicing tape,[32] the material is transferred to a highly doped silicon substrate with 500 nm thermally grown surface oxide and pre-defined grid markers, and then imaged using an optical microscope, see Figure 1c. After selection of quasi-1D structures, electron beam lithography is used to define the contacts, Figure 1d and Figure 1e. As contact material, following Shen et al.,[38] we use a bismuth layer with thickness between 25 and 45 nm, capped with 100 nm gold. For most devices, in particular devices B, C, D, the contacts were deposited in a standard evaporation chamber equipped with an electron-beam evaporator for Bi and Au, while device A was prepared in a molecular beam epitaxy (MBE) system in which Bi and Au are evaporated from Knudsen cells; see the Supporting Information for detailed device and fabrication information. Lift-off was subsequently carried out in hot acetone.

As already discussed in literature,[36] some tubular MoS$_2$ nanostructures collapse during growth, forming flat, sometimes twisted multilayer MoS$_2$ nanoribbons without broken bonds at their edges. Figure 1f displays a scanning electron microscopy (SEM) image of one of our devices. From its shape, also confirmed by atomic force microscopy (AFM), the MoS$_2$ structure can be characterized as such a nanoribbon.

## 3. Room-Temperature Characterization

To place the fabrication results using bismuth into context, **Figure 2**a compares the room temperature resistances of our devices to previously fabricated ones using different contact metals. As already mentioned, for metallic contacts, a low work function is expected to be advantageous. Consequently, we here select titanium[40,41] ($\Phi_{Ti} = 4.3$ eV) as a previously used material, and hafnium[40] ($\Phi_{Hf} = 3.9$ eV) as an experimental alternative. In addition, inspired by Liu et al.,[42] we test thin copper films ($\Phi_{Cu} = 4.6$ eV). Each point in Figure 2a corresponds to the room temperature resistance of one measured MoS$_2$ nanoribbon or nanotube segment at $V_g = 0$.

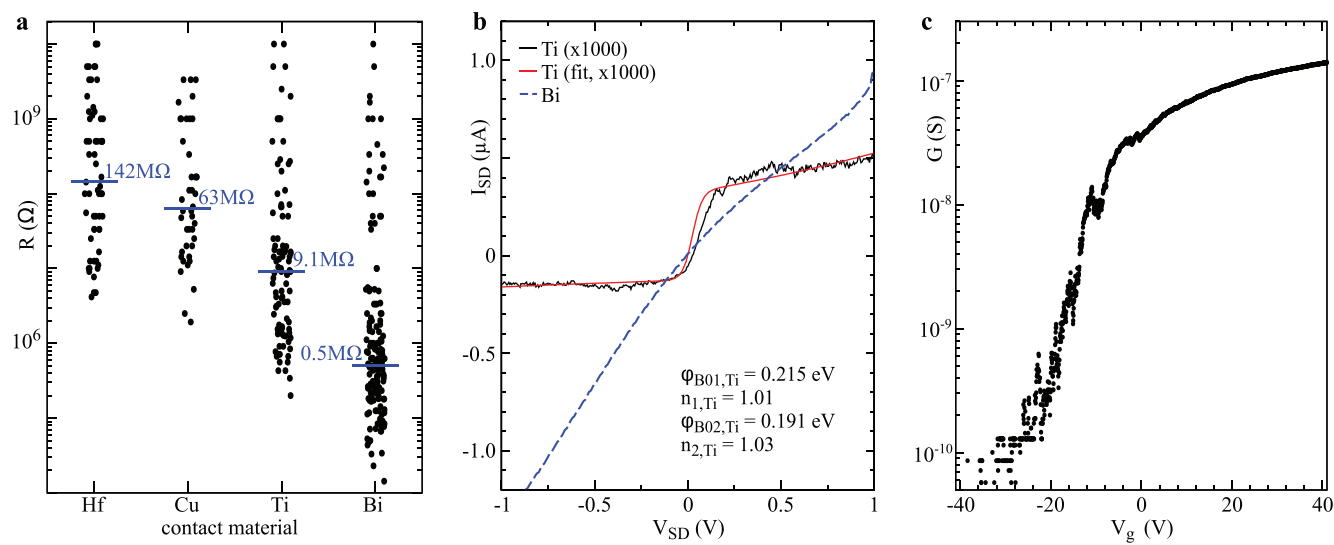

**Figure 2.** Room-temperature measurements. a) Comparison of bismuth contacts with previously tested contact materials hafnium, copper,[42] and titanium; each point represents the two-terminal room temperature resistance of a MoS$_2$ device. The blue horizontal lines mark the median value for each material. b) Comparison of the source–drain current $I(V_{SD})$ of a titanium-based and a bismuth-based device, demonstrating much improved linearity when using bismuth. The fit of the titanium curve is based on a double Schottky barrier model by Grillo et al., with $\varphi$ denoting the barrier heights and $n$ denoting the ideality factors of the barriers.[43] c) Low-bias differential conductance $G(V_g)$ as function of gate voltage, demonstrating semiconducting behavior of a nanoribbon.

For all materials, a wide scattering of the resistance values is observed. Nevertheless, the plot with its logarithmic resistance scale clearly demonstrates the merits of bismuth-based contacts, with a reduction of the median two-point resistance from 9.1 MΩ in the case of titanium to 0.5 MΩ by more than one order of magnitude and minimal resistances down to 50 kΩ.

Figure 2b compares the room-temperature source–drain characteristics $I(V_{SD})$ of a titanium-based and a bismuth-based device. The titanium contact based device displays strongly nonlinear behavior consistent with two Schottky barriers back-to-back in series.[43–45] The red solid line in Figure 2b is a fit using the double barrier model of Grillo et al.[43] with the resulting fit parameters given in the figure. The fit clearly describes the data well, with a barrier height of $\Phi \approx 0.2$ eV on both sides.

The blue dashed line, displaying measurement data of a bismuth-contacted device, is considerably more linear, with a much higher current (note the scaling factor of 1000 for the titanium device). Only a very weak single-barrier diode-like behavior is still observed,[46] with a larger forward current for negative $V_{SD}$ and a smaller reverse current for positive $V_{SD}$. Here, the Schottky barrier at one contact is negligible; the second contact dominates conduction but is still very transparent.

The gate voltage dependence of a similar bismuth-contacted device on the same chip is shown in Figure 2c. A clear transistor-like behavior of the semiconducting nanodevice can be observed, with the conductance reaching the resolution limit of our lock-in amplifier below $V_g = -20$ V. About half of the devices characterized in detail display a qualitatively similar semiconducting behavior; for the remaining ones, no clear gate voltage dependence of the conductance can be observed within the tested region (see also Table S1, Supporting Information).

From theory, all MoS$_2$ nanotube walls are expected to be semiconducting.[47] For multiwall carbon nanotubes, a considerable tunnel coupling between layers has been identified,[48,49] which can lead even to an additional level degeneracy.[50]

Screening from a backgate, however, is for MoS$_2$ 15 times larger than for graphene, and the resistance between layers 20 times larger than for graphene.[51] This leads to a complex electronic behavior of gated multilayers.[52] For nanoribbons, a detailed model additionally has to take into account the connection between upper and lower layers and bending at the ribbon edge. So far, we have been unable to identify systematic differences in behavior between nanoribbons and nanotubes. In either case, differing cutoff voltages of different walls seem plausible.

## 4. Low-Temperature Measurements

**Figure 3** displays overview measurements of two bismuth contacted devices, at base temperature $T \simeq 15$ mK of a dilution refrigerator in a standard Coulomb blockade measurement setup. The absolute value of the dc current $|I(V_g, V_{SD})|$ is plotted as function of applied gate voltage $V_g$ and bias voltage $V_{SD}$ in logarithmic color scale.

In the measurement of device A shown in Figure 3a, a clear transition is visible. For $V_g \geq -19$ V the device is highly transparent. While Coulomb blockade regions are still visible their characteristic charging energy is on the order of $E_C \simeq 0.2$ meV, see the inset of Figure 3a. The tips of the single electron tunneling regions mostly touch, indicating negligible voltage drops outside this central quantum dot. Below $V_g = -19$ V, larger structures resembling additional Coulomb blockade regions become visible, leading to the emergence of an energy gap. This behavior is consistent with a semiconducting system, where an electronic potential well breaks up into several puddles close to the band edge.

A different outcome is observed in device B, a nanotube with approximate diameter of 100 nm, as shown in Figure 3b. It displays irregular Coulomb blockage over a large gate voltage range without systematic changes. The total investigated gate voltage interval was significantly larger than the representative region of Figure 3b, covering $-80$ V $\leq V_g \leq 24$ V. The charging energy, read out from Coulomb blockade, is with up to $E_C \simeq 1$ meV larger than for device A, indicating a smaller quantum dot. This is consistent with the smaller dimensions of the nanostructure, including the channel length of $L = 150$ nm compared to $L = 270$ nm in device A (see **Table 1**).

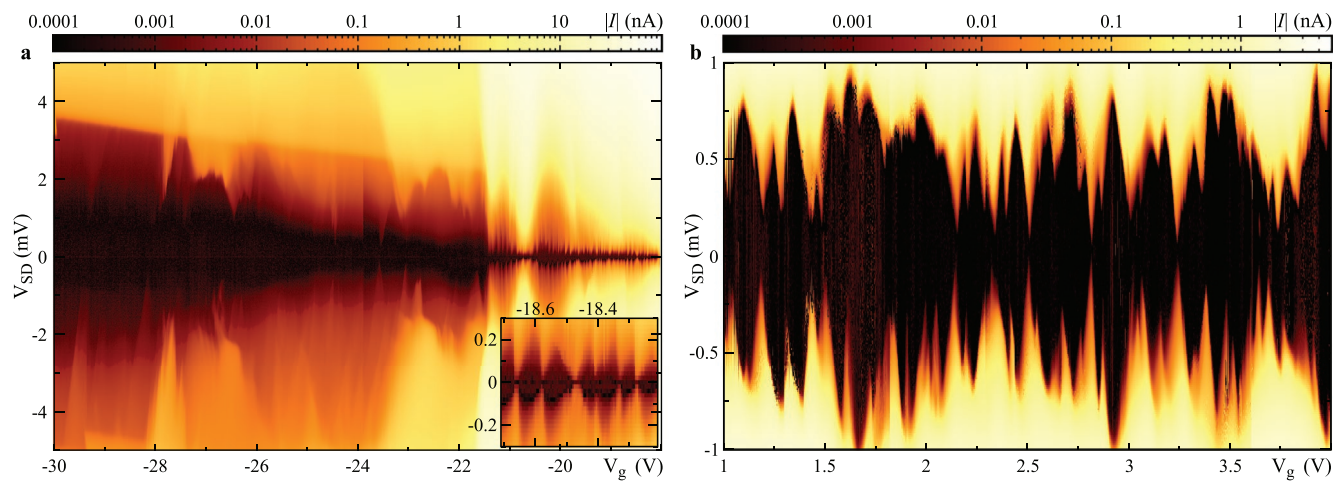

**Figure 3.** Low-temperature transport measurements ($T \simeq 15$ mK) of two devices; absolute value of the dc current $|I(V_g,V_{SD})|$ in logarithmic color scale, as function of back gate voltage $V_g$ and bias voltage $V_{SD}$. a) Device A displays a transition into a band gap below $V_g = -21.5$ V. Coulomb blockade phenomena of several characteristic energy scales are visible. Inset: Detail zoom of the low-bias region. b) Typical measurement of device B, showing disordered Coulomb diamonds of varying sizes. No characteristic large-scale variation in $V_g$ is observed for $-80$ V $\leq V_g \leq 24$ V.

Table 1. Structural and electronic parameters of the devices: segment/channel length $L$, width $w$, height $h$ from AFM measurement, characterization as nanotube (T) or nanoribbon (R), room temperature resistance $R_{RT}$ at $V_g = 0$, observation of a bandgap, typical low-temperature charging energy $E_C$, and gate capacitance $C_g$ extracted from Coulomb blockade. An extended version of this table can be found in Table S1 (Supporting Information).

| Device | $L$ [nm] | $w$ [nm] | $h$ [nm] | T/R | $R_{RT}$ [kΩ] | gap | $E_C$ [meV] | $C_g$ [aF] |
|---|---|---|---|---|---|---|---|---|
| A | 270 | 500 | 18 | R | 97 | yes | 0.25 | 2.0 |
| B | 150 | 90 | 115 | T | 43 | no | 1.1 | 0.72 |
| C | 200 | 155 | 20 | R | 250 | no? | 1.85 | 0.46 |
| D | 220 | 360 | 20 | R | 178 | yes | 2.1 | 1.6 |

**Figure 4** shows detail stability diagrams for two additional MoS$_2$ nanoribbons where the single electron tunneling regions of finite current nearly reach the $V_{SD} = 0$ line (indicating no external voltage drop). The behavior of device C, Figure 4a, is similar to that of device B, with no systematic large-scale effect of an applied gate voltage, though here only the region $-10\,V \leq V_g \leq 10\,V$ was tested. Device D, Figure 4b, displays a transition into a band gap for negative $V_g$. Typical charging energies of $E_C \simeq 1.85$ meV and $E_C \simeq 2$ meV are observed, see Table 1. While disorder and trap-state behavior is present, full transport blockade only takes place at very low energy scales ≲ 0.5 meV.

For nanoribbons, open questions remain regarding the precise electronic device geometry. Gate capacitances $C_g$ in carbon nanotube devices are often modeled analogous to a thin wire above a metallic plane.[53] Our MoS$_2$ nanoribbon devices, however, more resemble plate capacitors, with the area given by channel length $L$ and ribbon width $w$. Our experimental $C_g$ values lie between both approximations, see Table S1 (Supporting Information). Deposited electrodes predominantly contact the outermost wall of a multiwall object; a speculative explanation for the capacitance values is the formation of quasi-1D channels in the curved regions at the nanoribbon edge, due to bending and strain.[9,10,54–56]

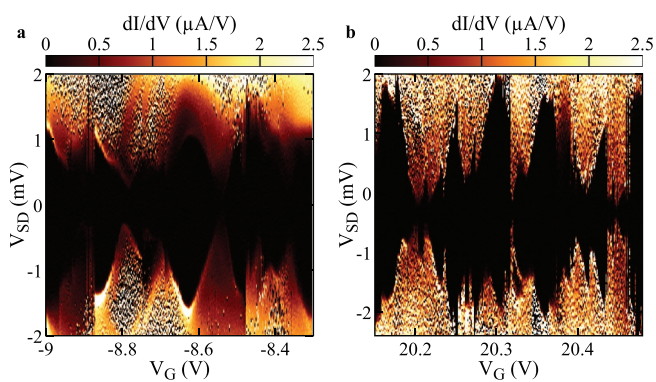

**Figure 4.** a,b) Detailed stability diagrams of device C (a) and device D (b): numerical derivative of the dc current $dI/dV_{SD}(V_g,V_{SD})$, plotted against gate voltage $V_g$ and bias voltage $V_{SD}$. In both cases, disordered Coulomb blockade is visible. The single electron tunneling regions nearly reach $V_{SD} = 0$, indicating low voltage drop outside the main quantum dot.

## 5. Discussion—Toward Quantum Devices

Compared to the state of the art of carbon nanotube quantum dots and devices,[57] significant technological improvements on MoS$_2$ nanotubes are still required. Broadly, these improvements fall into three categories. The first one—transparent and non-destructive contacts—is addressed here. With conductances approaching $e^2/h \simeq 39$ μS, see, e.g., Figure 4, and with the reduction of charge traps below the contacts, this objective is within reach.

The second category of improvements is the reduction of disorder within the device—achieving clean, repetitive Coulomb oscillations of conductance with only one charging energy scale corresponding to the entire active region. Amorphous silicon oxide surfaces are known to harbor charge traps,[58,59] leading to random, possibly time-dependent potentials. Using a substrate with less surface states, as, e.g., hexagonal boron nitride,[60] and shielding the influence of underlying layers[61,62] are obvious next steps here. Alternatively, MoS$_2$ nanotubes or nanoribbons can be suspended;[63–65] for nanotubes this would also help maintain the rotational symmetry.

The third category is the reduction of the device size—toward quantum confinement and discrete electronic states. No indications of a repetitive shell structure have been observed in this work. Scaling conditions for achieving such a spectrum with clear quantum effects are given by the comparatively large effective electron mass in the conduction band, expected to be $m_e^* \simeq 0.45\,m_e$.[9–11] However, the temperature evolution of several Coulomb oscillations of device A (see also the inset of Figure 3a) is shown in **Figure 5**, in the parameter region where only small-scale Coulomb blockade regions exist in an otherwise transparent device, and reveals a highly interesting detail.

As can be seen from Figure 5c, all five evaluated peaks shown in Figure 5a,b are temperature-broadened, with the full width at half maximum (FWHM) increasing linearly with $T$. The expected FWHM from only thermal broadening in the leads, $\Delta V_g = 3.52\,k_B T/e\alpha$, taking into account the gate conversion factor $\alpha = 0.0031$, is given in the figure as line for comparison.[66] Below $T = 100$ mK, corresponding to an energy scale $k_B T = 8.6$ μeV, all five current peaks clearly decrease in height with increasing temperature. This indicates that we have approached the regime of transport through a single quantum level,[66] where the maximum current scales as $I_{max} \propto T^{-1}$. At higher temperature, the peak amplitudes become constant and later increase again, meaning that the condition $k_B T \ll \Delta\varepsilon$ is not fulfilled for the level spacing $\Delta\varepsilon$ anymore.

A simple estimation for a hard-wall 1D quantum box with $m_e^* = 0.45\,m_e$ and length $L$ leads to quantization energies $\varepsilon(n,L) =$ $(n^2 h^2)/(8mL^2) = 8.36 \times 10^{-19}$ eV m$^2 \times n^2 L^{-2}$ and $\Delta\varepsilon(0.3\ \mu m) = \varepsilon(2, 0.3\ \mu m) - \varepsilon(1, 0.3\ \mu m) = 28\ \mu eV$, consistent with the thermal energy scale discussed above. Longitudinal confinement at a scale of $L = 0.1\ \mu m$ along a MoS$_2$ nanotube is possible using local gate electrodes and shall then allow for considerably larger confinement effects, e.g., $\Delta\varepsilon(0.1\ \mu m) = 0.25$ meV in the box model.

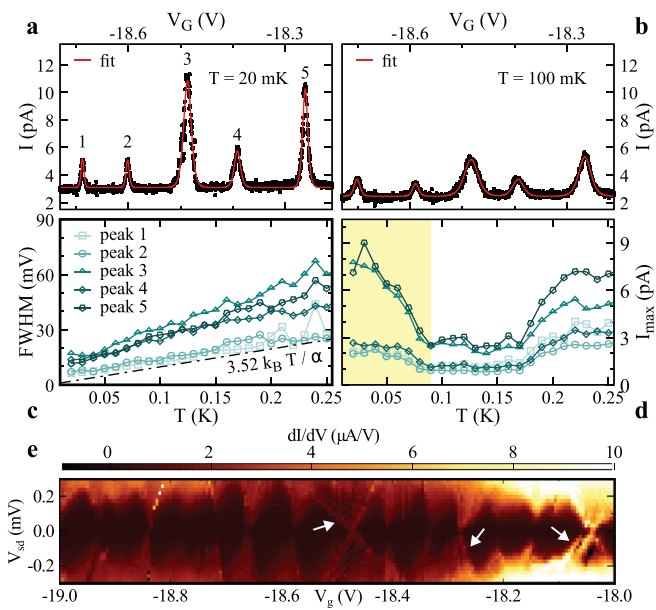

**Figure 5.** a,b) Nanoribbon device A: sequence of Coulomb oscillations of the dc current in linear response, measured at 20 mK (a) and 100 mK (b), together with a fit (red line) based on five current peaks each modeled as $I(V_g) = I_{max} \cosh^{-2}((V_g - V_g^0)/\sigma)$ and a constant background. c,d) Fit parameters peak full width at half maximum (FWHM) (c) and peak height (d) as function of temperature. e) Numerically derived differential conductance $dI/dV_{SD}(V_g, V_{SD})$ of the region containing the Coulomb oscillations of (a–e). Due to potential drift, an unambiguous identification of the peaks is not possible. Larger versions of the plot with different color scales can be found in the Supporting Information.

Figure 5e shows the differential conductance in the parameter region containing the five peaks of Figure 5a–d. Larger versions of the plot can be found in Figure S6 (Supporting Information). Due to drift of the electrostatic potential between measurements, an unambiguous identification of the peaks is not possible. Despite the enlargement being at the edge of our measurement resolution, discrete conductance lines seem to appear at several points, see the white arrows in the figure. While this is in agreement with above conclusions, the conductance data of Figure 5e is not detailed enough yet to allow further statements or evaluation.

## 6. Conclusion

Bismuth as contact layer has improved the transport properties of MoS$_2$ nanoribbon and nanotube devices significantly. Room temperature resistances on the order of 100 kΩ are now regularly observed, and devices have been shown to remain well-conducting even at dilution refrigerator base temperature. We have observed fully or nearly fully "closing" Coulomb diamonds, indicating the predominant absence of sequential charge traps at the contacts. Measurements indicate single quantum level transport at temperatures below 100 mK; with this, we reach beyond classical Coulomb blockade into the regime of quantum dots and quantum confinement.[66]

The research focus now shifts to the subsequent challenges: suppressing random potentials and decreasing device sizes to allow discrete level spectroscopy. Past development of carbon nanotube devices provides clear paths for the former objective—suspending the MoS$_2$ nanotubes,[63,64] or replacing the underlying silicon oxide with hexagonal boron nitride.[61,62] Regarding confinement, the selected MoS$_2$ growth method from vapor phase produces nanotubes down to diameters $d \simeq 20$ nm.[33] Preselection of such nanotubes in combination with narrow gate electrodes is here the next step.

The promises of the MoS$_2$ nanotube material are manifold. Aside intriguing mechanical properties,[67–69] strong spin–orbit interaction in a non-centrosymmetric system leads to spin-split bands.[70] The threefold valley degeneracy has inspired theoretical comparisons of quantum dots to strong interaction phenomena of particle physics.[71] Beyond that, MoS$_2$ is at sufficient doping an intrinsic superconductor;[7,8] while not much specific theoretical work exists yet, the combination of geometry, spin–orbit interaction, and superconductivity may eventually lead to a novel approach toward topology-based phenomena and Majorana zero modes.[72,73]

## Supporting Information

Supporting Information is available from the Wiley Online Library or from the author.

## Acknowledgements

The authors thank the Deutsche Forschungsgemeinschaft for financial support via grants Hu 1808/4-1 (project id 438638106), Hu 1808/6-1 (project id 438640730), and SFB1277 (subprojects A01 & A08, project id 314695032), and the Slovenian Research Agency for financial support via grant P1-0099. The authors would like to thank S. Ludwig for insightful discussions and C. Strunk and D. Weiss for the use of experimental facilities. The data was recorded using Lab::Measurement.[74]

Open access funding enabled and organized by Projekt DEAL.

## Conflict of Interest

The authors declare no conflict of interest.

## Data Availability Statement

The data that support the findings of this study are available from the corresponding author upon reasonable request.